\documentclass[aps,pra,twocolumn,superscriptaddress,groupedaddress,amssymb,longbibliography]{revtex4}
\usepackage{graphicx}
\usepackage{centernot}
\usepackage{subfigure}
\usepackage{color}
\usepackage{dcolumn}
\usepackage{url}
\usepackage{amsmath}

\newcommand{\bc}{\begin{center}}
\newcommand{\ec}{\end{center}\hrule}

\newcommand{\half}{\frac{1}{2}}

\begin{document}
\title{The concept of free will as an infinite metatheoretic
recursion}   \author{Hanaan  Hashim}   \affiliation{Sibylgaze,  Industry
  Intelligence  GmbH  (Switzerland),  Bangalore,  India.}   \author{R.
  Srikanth}   \affiliation{Poornaprajna    Institute   of   Scientific
  Research,          Sadashivnagar,          Bangalore,         India}
\email{srik@poornaprajna.org}

\begin{abstract}
It is argued that the concept of  free will, like the concept of truth
in formal languages, requires a separation between an object level and
a meta-level  for being consistently defined.   The Jamesian two-stage
model, which  deconstructs free will  into the causally  open ``free''
stage with its closure in the  ``will'' stage, is implicitly a move in
this direction.   However, to avoid  the dilemma of  determinism, free
will additionally requires an  infinite regress of causal meta-stages,
making free choice a hypertask.  We use this model to define free will
of the  rationalist-compatibilist type.   This is  shown to  provide a
natural three-way  distinction between quantum  indeterminism, freedom
and  free will,  applicable  respectively  to artificial  intelligence
(AI),  animal agents  and human  agents.  We  propose that  the causal
hierarchy  in  our   model  corresponds  to  a   hierarchy  of  Turing
uncomputability.   Possible neurobiological  and  behavioral tests  to
demonstrate free will experimentally  are suggested.  Ramifications of
the   model   for   physics,   evolutionary   biology,   neuroscience,
neuropathological medicine and moral philosophy are briefly outlined.
\end{abstract}

\maketitle

\section{Introduction}

Free will (FW) is a concept  in philosophy that refers to the putative
capacity of a  human agent to control her behavior  by choices made by
an act  of will,  on basis  of her  personal motives,  convictions and
intentions.  The  concept rests on  the belief that human  behavior is
not  fully  determined by  external  causes.   Moreover, her  motives,
convictions  and intentions  are  themselves not  determined by  fully
external   causes,  but   self-determined.    From   a  common   sense
perspective,  we  feel  we  are  free in  making  decisions.   Yet  it
continues to be debated, even after centuries of argumentation, how to
coherently  define  free   will  and  whether  it   exists  in  Nature
\cite{Pin04}. % \cite{tim,linda,Pin04,Pin09}.

FW can be regarded as freedom  from some constraint: exactly what that
constraint is, has  remained moot.  There are  two broad philosophical
positions  on  FW:  \textit{Incompatibilism},  which  holds  that  the
relevant constraint is  determinism, and \textit{Compatibilism}, which
holds that determinism  is irrelevant to the definition  of free will,
and   that  determinism   and  FW   are  compatible.    

Two divergent Incompatibilist positions are \textit{Hard Determinism},
which regards FW as false  and determinism as true, and (metaphysical)
\textit{Libertarianism}, which regards determinism  as false and FW as
true.  A  FW \textit{Skeptic} is  an Incompatibilist who  goes farther
than a Hard Determinist and denies that FW is even a coherent concept.
From the  Skeptic perspective, indeterminism  no more allows  an agent
control  and self-determination  over actions  than does  determinism.
Since  determinism  and  indeterminism  are  the  only  two  logically
possible causal primitives, this view holds that very concept of FW is
meaningless.  This  Skeptic stand is  sometimes called the  dilemma of
determinism or the standard modern argument against FW.

The libertarian may imagine that housed  in her brain is an immaterial
agency  (such as  a soul  or homunculus)  that somehow  transcends the
cause  and effect  law  that holds  elsewhere in  Nature.   This is  a
logically tenable defense of  Libertarianism against Hard Determinism,
even if  an adherent of  the latter would  deny the existence  of such
immaterial  agencies.   But even  the  soul  offers Libertarianism  no
protection  against  the  Skeptic,   because  if  the  soul's  choices
transcend  causality,  they  must  be random,  or  results  of  random
properties, again making the case for FW bleak!

The present  work hopes  to convince  the FW  Skeptic that  a coherent
metaphysical account of  FW is possible. Of importance to  our work is
using an idea similar to that used  by Tarski to define the concept of
truth, except that here it is  applied to causality.  We think that FW
is   a  causal   primitive   different  from   both  determinism   and
indeterminism in being \textit{metatheoretic}.   Our attempt to define
it shows it to be a form of causation that straddles endless levels of
causality,  \textit{provided} we  wish to  form a  coherent scientific
narrative that fits  in with out intuitive sense of  self, freedom and
responsibility.

This  article is  structured as  follows.  A  version of  the Jamesian
two-stage model  is proposed in  Section \ref{sec:2S}, which  would be
useful for  our further  discussion.  Although such  a model  has been
considered as the defense of FW against the dilemma of determinism, we
point out in Section \ref{sec:rdilemma}  that the model fails, because
the   dilemma   can   be    recursively   resurrected.    In   Section
\ref{sec:RDoD},  we  show  that  an  infinite  recursion  of  Jamesian
two-stage models  restores a measure of  protection for libertarianism
against   the  dilemma.    This   model  is   adapted   to  define   a
Rationalist-Compatibilist FW.  The connection of FW to uncomputability
\cite{md1958} and Tarskian undefinability  for formal truth \cite{T44}
are  outlined  in  Section  \ref{sec:turing}.   We  then  discuss  the
consequences   of    the   model    for   neurobiology    in   Section
\ref{sec:noorba}, before concluding in Section \ref{sec:conclu}.

\section{The two-stage model, $2S$: A new hope \label{sec:2S}}

The two-stage model \cite{H09}, introduced  in its original form by W.
James  \cite{J56},  and proposed  in  various  forms  by a  number  of
researchers, is  intended to defeat  the dilemma of  determinism.  The
model posits that FW is a two-stage process: first there is freedom at
the lower  stage, and then  there is will  at the higher  stage, which
makes  a choice.   We present  below a  version of  it, which  we call
`$2S$',  with  several  changes  in   the  details,  as  described  in
subsections \ref{sec:l0} and \ref{sec:l1}.

\subsection{Physical stage $L_0$    (Freedom) \label{sec:l0}}    

At the moment  an agent's attention is drawn to  a conflict situation,
during a short  time span, called the \textit{selection  window}, in a
localized region of brain, which  we refer to as the \textit{free-will
  oracle}, probably in the pre-frontal cortex, the physical laws $W_0$
are put  on a ``causally  open'' mode.   In preparation of  entry into
this  mode,  alternative  options   $x_0$,  described  by  probability
distribution $P_0$, are generated in  the agent's brain.  The physical
laws $W_0$ only determine $P_0$, but  do not entirely fix the eventual
choice $x_0$.  And  the selection is not completely  determined by the
past history of  the \textit{physical} universe.  At the  close of the
selection  window, $W_0$  in FW  oracle  is re-set  to the  ``causally
closed''  mode, and  the  choice $x_0$  that is  available  on the  FW
oracle's  ``register'' at  that  moment is  expressed  as the  agent's
action.

If the agent makes the choice mechanically, without a focused exertion
of her will, then there is  no mental causation influencing the choice
(as described  below), and  at the  close of  the selection  window, a
random $x_0$ is selected according to probability distribution $P_0$.

\subsection{Metaphysical stage $L_1$  (Will) \label{sec:l1}} 

If the agent  decides to exert her  will, then her choice  must not be
random  but  instead  reflect  the  desires,  intentions  and  beliefs
characteristic  of   the  agent.   These  properties   constitute  her
\textit{cognitive   private  space},   and   may  not   have  a   full
representation in  the physical level  $L_0$.  Thus the  will produces
generally a \textit{deviation} from $P_0$.

A simple way to describe this situation mathematically is by:
\begin{equation}
x_0 = \lambda_1(P_0),
\label{eq:2S}
\end{equation}
where $\lambda_1$ is the ``will function'' that encodes the properties
of her  cognitive private  space. Function  $\lambda_1$ takes  note of
$P_0$, but  is \textit{not} part  of $W_0$,  because if it  were, then
there would be causal closure in the physical, and hence no freedom at
$L_0$. Instead,  $\lambda_1$ is  determined by  the laws  $W_1$, which
extend $W_0$ to $L_1$.

The  fact that  the causal  openness at  $W_0$ is  replaced by  causal
closure of $W_1$ in the consolidated system in $L_0+L_1$ suggests that
these two levels can be  ordered in a \textit{causal hierarchy}, which
we  represent  by  the  expression:  $L_1  \prec  L_0$.  The  ordering
expresses that $L_1$ causally precedes $L_0$.

\subsection{2$S$ and the dilemma of determinism}

In the absence of action by the  will, the FW oracle can be considered
as a \textit{probabilistic} input/output machine.  To underscore this,
sometimes  we   will  refer   to  such  indeterministic   behavior  as
\textit{freedom-without-will}.   We  introduce  a level  $L_\half$  as
essentially  $L_0$   equipped  with  a  source   of  pure  randomness.
Instances  of  freedom-without-will  have   their  causal  closure  in
$L_0+L_\half$,  but in  this case  there is  no deviation  from $P_0$.
When the will is exerted,  the random variable $X_0$ representing free
choice  deviates  from  $P_0$,   and  free  will  transcends  physical
causality.

The attempted defense  of FW against the dilemma  of determinism using
model $2S$ would be: FW is not deterministic because of freedom at the
physical  stage. Nor  is  it  random because  the  eventual choice  is
self-determined, being fixed by personal preferences via $\lambda_1$.

Though not  the definitive word on  FW, the model $2S$  brings the new
insight that  FW is  a new  \textit{causal primitive},  different from
determinism and indeterminism.  While the latter two can be associated
with a causally closed lawfulness, and thus defined on a single causal
level (say $L_0$), FW cannot. More of this below.

\subsection{Free will in $2S$ and formal truth \label{sec:resol}}

As we indicated in Ref.   \cite{NS14}, the 2$S$ feature of introducing
the metaphysical  level to define  FW on the physical  level parallels
Tarski's use  \cite{T44} of  a metalanguage  in order  to consistently
define the concept of arithmetic  truth in an object language.  Tarski
showed that  without careful separation  of the two levels,  one would
end  up  with  logical  antinomies  like  the  liar's  paradox  ``This
statement is false'' (which is true iff it is false).

Similarly, if we fail to separate  the ``free'' and ``will'' stages in
free choice,  we will  err in  reducing the  choice to  determinism or
randomness.  We believe  that part of the  difficulty in understanding
the nature  of FW  is due  to a  lack of  appreciation for  this level
separation.

With this  in mind,  we may  consider $L_0$ as  the object  stage, and
$L_1$ as the metastage. The free-willed agent with $2S$ structure will
be designated $\mathfrak{F}_1$. A deterministic physical system, which
is  causally closed  at  level $L_0$  is designated  $\mathfrak{F}_0$,
while  an indeterministic  quantum  physical system,  which has  $L_0$
freedom-without-will will be denoted $\mathfrak{F}_\half$.

\section{Dilemma of determinism revisited\label{sec:rdilemma}}

Although the  model $2S$ resolves  the dilemma of determinism  after a
fashion, still the dilemma can be  resurrected at level $L_1$.  To see
this,  note that  the  two-stage agent  $\mathfrak{F}_1$,  taken as  a
whole, is deterministic, in view  of Eq.  (\ref{eq:2S}), and must thus
lack  FW.  This  is  just Schopenhauer's  argument, who  picturesquely
said, ``Man can  do what he wants  but he cannot will  what he wants''
\cite{S39}, in  his prize-winning essay  in response to  the challenge
posed by the Royal Norwegian Academy of Sciences in 1839 \cite{N39}.

To reinstate free  will, we allow the will function  $\lambda_1$ to be
freely chosen,  i.e., we apply  $2S$ to the selection  of $\lambda_1$.
We  extend  the  FW  oracle  from the  physical  level  $L_0$  to  the
metaphysical  $L_1$.   Within the  selection  window,  the laws  $W_1$
governing  level $L_0+L_1$  become  momentarily causally  open in  the
region of the FW oracle.  We thus have \textit{freedom of the will} at
$L_1$ whereby  the $L_0+L_1$ (i.e., physical-metaphysical)  history of
the universe does not fix $\lambda_1$.

We introduce  a metalevel  $L_2$, where  a higher-order  will function
$\lambda_2$  (``will   free  will''),  provides  causal   closure,  by
deterministically selecting a particular  $\lambda_1$. We can think of
$\lambda_2$ as representing  a deeper aspect of  the agent's character
or  disposition and  as  determining  the type  of  the will  function
$\lambda_1$ that she  will select.  Function $\lambda_2$  comes from a
further interior  layer of the  cognitive private space of  the agent.
By adding an element of spontaneity and self-determination, it thwarts
the selection  of $\lambda_1$ from  being modelled as  an input/output
process at level $L_1$.  In  the metatheoretic representation, this is
tantamount to treating  $\mathfrak{F}_1$ as the object  system, and an
$L_2$-aspect  as  the  metasystem.    A  simple  mathematical  way  to
represent this is by:
\begin{equation}
\lambda_1 = \lambda_2(P_1),
\label{eq:2S+}
\end{equation}
where    $P_1$   is    a   \textit{probability    function}   encoding
$L_1$-preferences according  to laws $W_1$.  If  the higher-order will
$\lambda_2$   is  not   exerted,   then  $P_1$   would  describe   the
indeterministic  selection of  will $\lambda_1$.   But if  this ``will
free will''  is exerted, then there  will be deviations from  $P_1$ in
the selection of $\lambda_1$.

Substituting Eq.  (\ref{eq:2S+}) into Eq. (\ref{eq:2S}), we obtain:
\begin{equation}
x_0 = \lambda_1(P_0) =  [\lambda_2(P_1)](P_0)
\equiv \lambda^\ast_2(P_1,P_0).
\label{eq:2S++}
\end{equation}
The  second-order   will  $\lambda_2$   selects  a   first-order  will
$\lambda_1$ depending  on $P_1$,  and then $\lambda_1$  selects $x_0$.
The extended  causal hierarchy is  $L_2 \prec L_1 \prec  L_0$, whereby
$L_2$ causally precedes $L_1$, which in turn precedes $L_0$.

Here precedence refers  not just to chronology but  to recursion depth
in  the agent's  cognitive private  space.  It  does not  refer to  an
\textit{earlier}   event   on   the   same  causal   level,   but   to
\textit{deeper} or  higher-order cause.   For example,  if $\lambda_1$
inclines  Alice to  give Bob  a gift,  then $\lambda_2$  could be  her
character  trait  of  wanting  to   help  him.   A  yet  deeper  cause
($\lambda_3$) would  be wanting for good  to him or people  she likes.
And so on.

We  denote by  $\mathfrak{F}_2$ the  type of  agent for  who there  is
causal closure  at $W_2$, and by  $\mathfrak{F}_{\frac{3}{2}}$ one for
who  the will  function $\lambda_1$  is selected  indeterministically.
Although  the existence  of $\mathfrak{F}_2$  resolves the  dilemma of
determinism on  $L_1$, still the  dilemma can be resurrected  at level
$L_2$.  This is evident seeing  that the action of $\lambda^\ast_2$ in
Eq.  (\ref{eq:2S++}) is deterministic.

To reinstate free  will at $L_2$, we recursively apply  $2S$.  We make
$L_2$-law  $W_2$  at the  FW  oracle  also  causally open  within  the
selection  window. Let  the  freedom of  $\lambda_2$  be described  by
probability  function  $P_2$,  determined by  $W_2$.   Causal  closure
occurs at  $W_3$ via higher-order  FW $\lambda_3$ from an  even deeper
aspect ($L_3$-aspect) of the agent's cognitive private space. We have
\begin{equation}
\lambda_2 = \lambda_3(P_2),
\label{eq:l2}
\end{equation}
Substituting   Eq.     (\ref{eq:l2})   into
Eq. (\ref{eq:2S+}), we obtain:
\begin{eqnarray}
x_0 &=& [\lambda_2(P_1)](P_0) =
[[\lambda_3(P_2)](P_1)](P_0) \nonumber \\
&\equiv& \lambda^\ast_3(P_2,P_1,P_0).
\label{eq:w2}
\end{eqnarray}
We extend the causal hierarchy as $L_3 \prec L_2 \prec L_1 \prec L_0$.
An  agent with  causal closure  of her  choosing process  in $W_3$  is
denoted $\mathfrak{F}_3$  and one  with freedom-without-will  on $L_2$
(i.e., $L_2$-indeterministic) is denoted $\mathfrak{F}_{\frac{5}{2}}$.

Although  the existence  of  $\mathfrak{F}_3$ appears  to resolve  the
dilemma  of determinism  on stage  $L_2$, still  the  dilemma can  be
resurrected  at stage  $L_3$ since  $\mathfrak{F}_3$, in  view of  Eq.
(\ref{eq:w2}), can be considered as  a deterministic system.  This may
be  seen  as  a   further  higher-order  extension  of  Schopenhauer's
argument.

To prevent the  dilemma of determinism at stage $L_3$,  we introduce a
$2S$ model on  top of this stage, with $\mathfrak{F}_3$  as the object
system,  and  $L_4$   as  the  metastage.   But   then  the  resultant
$\mathfrak{F}_4$ will be deterministic.  We  require $L_5$ to obtain a
$2S$ model for $\mathfrak{F}_4$ to avoid the dilemma of determinism at
level $L_4$. Continuing this trend indefinitely, at level $L_n$, where
$n$ is any positive integer, we have:
\begin{equation}
\lambda_{n-1} = \lambda_n\left(P_{n-1}\right).
\label{eq:lambn}
\end{equation}
Substituting    this    recursively    into    lower    levels    into
Eq. (\ref{eq:2S++}), we obtain:
\begin{eqnarray}
x_0 &=&  [[\cdots\left[[\lambda_n\left(P_{n-1}\right)]\left(P_{n-2}\right)\right]\cdots](P_1)](P_0),\nonumber \\ %\label{eq:lamna}\\
 &\equiv& \lambda^\ast_{n}\left(P_{n-1},\cdots,P_0\right).
\label{eq:lamn}
\end{eqnarray}
Evidently, we can still resurrect  the dilemma of determinism at level
$L_{n+1}$   no  matter   how  large   $n$  is,   since  the   fact  of
$\lambda_n^\ast$ in  Eq. (\ref{eq:lamn}) is a  deterministic function.
Thus, the problem  posed by this dilemma does not  disappear for agent
$\mathfrak{F}_n$ but is merely postponed.

In response to  this seemingly insurmountable difficulty  posed by the
recursive version  of the dilemma  of determinism, the only  option to
save  libertarianism seems  to be  to let  a free-willed  agent be  an
infinite-stage  entity,  $\mathfrak{F}_\infty$.   What this  means  is
that, for any  \textit{finite} integer $n$, the laws  ($W_n$) of cause
and effect at  the stage $L_n$ will lack causal  closure in the region
of  the  extended FW  oracle  during  the  selection window,  and  the
selection of will $\lambda_n$ cannot be modelled as a probabilistic or
deterministic input/output  system at  stage.  The causal  closure for
$W_n$ will come through a higher-order cause $\lambda_{n+1}$, which is
determined  by the  $L_{n+1}$-aspect of  the cognitive  private space.
Therefore, the  agent's choice  of $\lambda_n$ will  transcend $n^{\rm
  th}$-order causality $W_n$, so that the choice of $\lambda_n$ should
be regarded as spontaneous and self-determined at that stage.

In FW  so understood, there is  libertarian freedom in the  sense that
the  agent's free  choice has  an  inexhaustible causal  depth in  her
cognitive private space. Perhaps, when we  scan the inner space of our
consciousness,  and  feel  that  our  choices are  free,  it  is  this
infinitude that we  grasp intuitively, and feel inclined  to report as
genuine personal  freedom.  

Identifying human agents with $\mathfrak{F}_\infty$ also means that FW
is at least a \textit{supertask} \cite{Lar09}, a process that involves
a sequence  of countably infinite  number of steps executed  in finite
time.   In  Section  \ref{sec:turing},  we will  present  an  argument
suggesting  that free  choice is  probably even  a \textit{hypertask},
which involves uncountably many steps executed in finite time.

\section{Free will as an infinite metatheoretic recursion
\label{sec:RDoD}}

The free agent $\mathfrak{F}_\infty$ in our model is an infinite-stage
entity   straddling  the   physical   $L_0$  and   the  ``final''   or
``infinite-th''  stage, denoted  $L_\infty$. For  simplicity, we  will
refer to  the agents $L_0$-aspect  as the ``physical aspect''  and the
$L_\infty$-aspect as ``transfinite aspect''.  In a conflict situation,
a response is  initiated at the transfinite aspect  and transmitted to
the  physical aspect,  where it  manifests  as the  choice $x_0$.   It
stands to reason that the  final destination to which information about
the sensory input is taken,  before the agent's response is initiated,
must also be the transfinite aspect.

\subsection{An infinite, staged causation\label{sec:stagec}}

Extending (\ref{eq:lamn}), we can represent the choice $x_0$ through a
sequence  of  downward  causations  starting  from  the  ``transfinite
preference'' $P_\infty$:
\begin{eqnarray}
x_0 &=&  [[[\cdots[[[\lambda_\infty(P_\infty)]]\cdots]](P_{2})](P_1)](P_0)
\nonumber\\
&\equiv& \lambda^\#_n[P_n,P_{n-1},\cdots,P_\infty](P_{n-1},\cdots,P_0).
\label{eq:x0}
\end{eqnarray}
The interpretation is that $\lambda_n^\#$, the will at stage $L_n$, is
fixed by higher-order  preferences, and then selects  an outcome $x_0$
depending on lower-order preferences.

The form of Eq.  (\ref{eq:x0})  suggests that the larger the recursion
depth $n$,  the fewer the higher-order  preferences $P_{n+1}, P_{n+2},
\cdots$ that could sway $\lambda_n^\#$  from the motivation encoded by
$P_\infty$.  In Eq. (\ref{eq:x0}), suppose that $0_{n-1}, \cdots, 0_0$
represent the  probability functions  $P_{n-1}, \cdots, P_0$  that are
\textit{unbiased} in the sense of being consistent with $P_\infty$.

Replacing the lower-order preferences by  their unbiased values in Eq.
(\ref{eq:x0}), we now define the $n^{\rm th}$-order \textit{intent}
\begin{equation}
x_n = \lambda^\#_n\left[P_n,P_{n+1},\cdots, P_\infty\right]
\left(0_{n-1},\cdots,0_0\right),
\label{eq:nintent}
\end{equation}
meaning that $x_n$ is the choice  $x_0$ that \textit{would} be made if
there are no distortions downwards  from level $L_{n-1}$.  Thus we may
call $x_\infty$ as the ``prime intent'' or ``transfinite intent'', the
option   that   would  be   selected   if   the  will   at   infinity,
$\lambda_\infty$, were to act unthwarted on the physical.

During the act of free choice, the prime intent is replaced stage-wise
by lower-order  intents, until  the final choice  is reached.   We may
refer to this infinite train
\begin{equation}
\textbf{X}  \equiv  x_\infty   \rightarrow  \cdots  x_n  \rightarrow
x_{n-1} \rightarrow \cdots \rightarrow x_1 \rightarrow x_0,
\label{eq:descent}
\end{equation}
as the ``descent of the will''.  This immediately evokes the notion of
FW  as  the effectiveness  of  communication  of $x_\infty$  from  the
transfinite aspect to the  physical aspect, undistorted by lower-order
preferences.  The will  is stronger, if this  channel of communication
(``volition   channel'')  is   clearer,  uncluttered   by  lower-order
motivations, beliefs  and desires inconsistent with  their transfinite
counterparts.   This   line  of  thought   forms  the  basis   of  the
compatibilist FW introduced below.

\subsection{Rationalist-compatibilist free will}
 
The model above can be extended to  protect FW from what may be called
the  `rationalist/robot paradox'.   By  definition,  a free,  rational
agent will, when faced with a  choice, select the optimal option.  His
behavior is  completely predictable, assuming  that there is  a single
rational option.  For a  libertarian, rationality appears to undermine
freedom  \cite{Pec09}.  Now  this  is not  the case,  as  viewed by  a
compatibilist.   But  the  rationalist/robot   paradox  asks  how  the
compatibilist  would differentiate  a rational  agent from  an optimal
robot programmed to choose rationally.

Following the  line of thought indicated  in Section \ref{sec:stagec},
we would like to think of the correlation between the prime intent and
final choice as a measure of FW, since it expresses how well the agent
is  able to  hold  on  to her  prime  intent  by overcoming  deviating
influences.    However,  this   correlation  would   stay  as   merely
incidental, unless the physical aspect holds $x_\infty$ as her purpose
or  motive.   For  this,  she  must  have  cognizance  of  $x_\infty$.
Precisely this defines the role played by the agent's rational faculty
or reasoning in  FW.  We will refer to  an $\mathfrak{F}_\infty$ agent
equipped with the rational faculty by $\mathfrak{F}_\infty^\sharp$.

The role played by reason is crucial.  Without it, the physical aspect
has no  motive to  deviate behavior from  that determined  by physical
causality  $W_0$.   Now there  may  be  random deviations  from  $P_0$
(applicable  to  animal agents),  but  they  would  be devoid  of  any
systematic or deliberate  attempt to transcend $W_0$.   By contrast, a
human agent,  on the recommendation  of reason, tries to  overcome the
imposition   of   $W_0$   by   trying   to   deviate   $X_0$   towards
$X_\infty$.  Here the  quantity $X_j$  represents the  random variable
corresponding to  $x_j$, i.e., values  of variables associated  with a
probability distribution.   Thus the  reasoning faculty serves  as the
basis through  which the  opportunity provided  by causal  openness is
exploited.

An  agent is  free to  the  extent that  she  is able  to enforce  her
transfinite will on her  physical choice.  (Complications arising from
the corruption  of the rational  faculty will be ignored  here.)  This
gives us a quantification of Rationalist-compatibilist FW:
\begin{equation}
\mathfrak{f} = \textrm{Corr}(X_\infty:X_0),
\label{eq:cFW}
\end{equation}
where Corr is any measure normalized  so that $-1 \le \mathfrak{f} \le
+1$.   The   rational   free-willed    agent   is   characterized   by
$\mathfrak{f}=1$, while a person completely under the sway of material
nature, by $\mathfrak{f}=-1$.

Lacking (substantial)  reasoning, an animal may  be represented simply
by $\mathfrak{F}_\infty$.  The animal is free, but not free-willed. We
express this insight with the expression:
\begin{equation}
\textrm{Freedom} + \textrm{reason} = \textrm{free~will}
\label{eq:RCeq}
\end{equation}
Quantum matter, or  in particular quantum AI, which  remains under the
scope of physical causality, is  a $\mathfrak{F}_\half$ agent, while a
classical robot  is a  $\mathfrak{F}_0$ agent.  It  is clear  how this
Rationalist-compatibilist account  protects FW  from rationalist/robot
paradox: a  deterministic robot is  a $\mathfrak{F}_0$ agent,  while a
rational  free-willed agent  is  a $\mathfrak{F}_\infty^\sharp$  agent
with $\mathfrak{f}=1$.

The brain  is arguably a  special organ, whose physical  structure has
somehow  been evolved  equipping  it  with a  FW  oracle, providing  a
gateway to  the transfinite  aspect.  AI lacks  this and  the physical
laws governing its dynamics are causally closed.

It  seems to  be an  interesting proposition  that plants  and ``lower
animals'' (like microbes), which lack  a central nervous system (CNS),
could be considered as intermediary  agents between quantum matter and
higher animals (like  mammals, reptiles and birds, which  have a CNS),
and thus represented by $\mathfrak{F}_K$,  where $\half < K < \infty$.
Some of these ideas are summarized in Table \ref{tab:animal}.

\begin{center}
\begin{table}
\begin{tabular}{l|c|c}
\hline
Entity & Agent type & Resource \\
\hline
Human & $\mathfrak{F}^\sharp_\infty$ & Free will (Freedom at  \\
   &  & all orders, plus reason)\\
\hline
Higher Animals &  $\mathfrak{F}_\infty$ & Freedom at all orders \\
(having a CNS) &    &   \\
\hline
Lower Animals &  $\mathfrak{F}_K $ & Freedom up to  \\
and plants  &  $(\half < K < \infty)$ &  a finite order    \\
(Lacking a CNS) &   &   \\
\hline
Quantum AI &  $\mathfrak{F}_\half$ &  First-order \\
   &  & freedom-without-will \\
\hline
Classical AI &  $\mathfrak{F}_0$ &  Determinism \\
\hline
\end{tabular}
\caption{Freedom gives  spontaneity, reason  gives self-determination,
  and freedom with  reason is free will.  AI, lacking  a FW oracle and
  being  thus  just a  special  configuration  of quantum  matter,  is
  described  by first-order  indeterminism.  By  contrast, the  higher
  animal or human brain, being equipped  with a FW oracle, has freedom
  at all orders.  Perhaps  the $\mathfrak{F}_\infty$ structure, common
  to humans and animals, is necessary for emotional behavior.}
\label{tab:animal}
\end{table}
\end{center}

\subsection{Causal vs Logical determinism}

The  resolution  of  the  rationalist/robot  paradox  shows  that  the
predictability  of  behavior does  not  imply  that the  behavior  was
causally  determined  (as  in  the robot's  case).  There  is  logical
determinateness  about the  rational agent's  behavior even  though he
transcends physical causality.  We express this idea by:
\begin{equation}
\textrm{Logical~determinism} 
\centernot\implies \textrm{Causal~determinism}.
\label{eq:LC}
\end{equation}
Now  this  result would  be  undermined  if  all humans  were  perfect
($\mathfrak{f}=1$) then, even if $P_\infty\ne P_0$, we would be led to
suspect  that  there  is  a  ``law  of  goodness'',  characterized  by
$P_\infty$,  that controls  human behavior.  However, some  people are
imperfect (having $\mathfrak{f}<1$), suggesting that human behavior in
general  transcends  causal determinism,  and  weakening  the need  to
undermine the above conclusion.

\subsection{Experimental test}

It   is  an   interesting  and   old  \cite{N39}   question:  how   to
experimentally demonstrate  the existence  of FW?  Our  model suggests
that  unfocussed or  casual  acts of  choice would  be  governed by  a
probability  distribution  $P_0$,  while  a  free-willed  action  with
deliberate  intent will  in  general produce  a  deviation from  $P_0$
towards $P_\infty$.

The observation of discrepancy  between the statistics of focussed and
unfocussed choice,  could be one  way to demonstrate the  existence of
FW.  Designing such an experiment may  not be easy, since the very act
of  focussing may  psychologically alter  $P_0$, so  that  an observed
deviation may  either be  due to will-induced  deviation or due  to an
alteration of $P_0$ or due to both.

\subsection{Why we are natural libertarians}

The model also helps make sense of people's instinctive inclination to
Libertarianism.  It  is  a  reasonable assumption  that  as  an  agent
introspectively scans the inner  space of her consciousness, depending
on how subtle  her awareness is, she can at  best objectively perceive
only so deep as there is freedom.

The  unfreedom and  higher-order causal  influences lying  beyond that
point become part  of her subjective consciousness, and  she is unable
to  consciously  experience  them,  though  she  may  deduce  them  by
observing her conscious choices and preferences.

% \cite{SBG}.

\section{Free choice and uncomputability \label{sec:turing}}

We now  explain a line of  thought indicated in Ref.   \cite{NS14}, on
the correspondence between  the causal hierarchy and  the hierarchy of
Turing  uncomputability,  in  reversed  ordering.  That  is,  given  a
$\mathfrak{F}^\sharp_\infty$  agent, $x_\infty$  is computable,  while
$x_m$ is harder to compute than $x_n$ if $m<n$.

The basic idea behind this claim  is the following.  Suppose one has a
computer program (or  Turing machine) so powerful that  it can compute
the  free  choice of  a  free-willed  agent,  using the  current  most
detailed description of  her brain state.  Now if  its prediction were
shown  to  the  agent,  being  free-willed,  she  may  contradict  the
prediction.  The conclusion  is that such a  powerful computer program
does not exist. We now consider a somewhat more detailed argument.

Given a  free-willed $\mathfrak{F}_\infty^\sharp$  agent $\textbf{A}$,
suppose  there is  a  computer $\mathcal{T}_C$,  programmable in  some
computer  language  $\mathcal{T}_L$  and  suitable  for  the  task  of
computing \textbf{A}'s  free choice  $x_0$.  Let $\hat{A}$  denote the
description of  \textbf{A} as  a computer program  in $\mathcal{T}_L$.
We assume  that all computer  programs that encode  in $\mathcal{T}_L$
the  description  of  free-willed  agents  are  denumerable  and  that
$\hat{A}$ is the  $a^{\rm th}$ program.  Similarly, one  is assumed to
be able to  encode situation any conflict situation  \textbf{J} in the
medium of  $\mathcal{T}_L$ by  a description $\hat{J}$,  and enumerate
them alphabetically as some number $j$.

If free  choice is computable, then the  computer $\mathcal{T}_C$ can,
given  the  enumerations for  the  $\mathcal{T}_L$-description of  the
agent and the  conflict situation, compute the agent's  free choice in
finite time, or:
\begin{equation}
\mathcal{T}_C(a; j) =
\left\{
\begin{array}{cc}
0 \Longleftrightarrow \textbf{A}(\textbf{J}) = 0 \\
1  \Longleftrightarrow \textbf{A}(\textbf{J}) = 1,
\end{array}, \right.
\label{eq:set}
\end{equation}
where for  simplicity we  have assumed the  outcome to  be two-valued.
(There is  no loss of generality,  since any computable output  can be
made  binary,  for  example  by  assigning ``0''  if  the  outcome  is
non-numerical or numerical and less  than 0, and ``1'' otherwise).  In
words,  the computer  produces  output 1  (resp.,  0) if  $\textbf{A}$
freely  chooses  1  (resp.,  0) when  faced  with  conflict  situation
\textbf{J}.

For any positive integer $j$, using this as a subroutine, we can build
another program:
\begin{equation}
\mathcal{T}_R(j) = \left\{
\begin{array}{cc}
1 \Longleftrightarrow \mathcal{T}_C(j;j) = 0 \\
0  \Longleftrightarrow \mathcal{T}_C(j;j) = 1,
\end{array} \right.
\label{eq:recur}
\end{equation}
which is  a representation  of the above  notion of  the uncooperative
free-willed agent.   Thus $\mathcal{T}_R$  outputs ``0'' on  input $j$
iff the  $j^{\rm th}$ free-willed  agent outputs ``1'' on  the $j^{\rm
  th}$ conflict situation.

We  can  now apply  the  computer  to  $\mathcal{T}_R$, so  that  from
Eqs. (\ref{eq:set}) and (\ref{eq:recur}):
\begin{equation}
\mathcal{T}_C(t_R; j) = \left\{
\begin{array}{cc}
1 \Longleftrightarrow \mathcal{T}_C(j;j) = 0 \\
0  \Longleftrightarrow \mathcal{T}_C(j;j) = 1,
\end{array} \right.
\label{eq:setrecur}
\end{equation}
where  $t_R$  is  the  enumeration  of  $\mathcal{T}_R$.   If  we  set
$j:=t_R$, then we  are led to a contradiction as  it would entail that
$\mathcal{T}_C(t_R,t_R) = 0$ iff $\mathcal{T}_C(t_R,t_R) = 1$.

To restore consistency, we infer  that $\mathcal{T}_C$ will never halt
on  inputs  $(t_R, t_R)$,  which  thereby  constitutes an  undecidable
G\"odel  sentence  under the  above  encoding.   We conclude  that  in
general $x_0$ will be uncomputable for the family of computer programs
considered.   

One   can   conceive    a   higher-level   ``meta-computer''   program
$\mathcal{T}_C^{(1)}$    that    is    able    to    decide    whether
$\mathcal{T}_R(t_R)$ equals 0 or 1, but  that is not contained in this
family.  If  the cardinality  of such  meta-computers is  greater than
$\aleph_0$  (countable infinity),  then  the  above diagonal  argument
based paradox can  be averted, because the meta-computers  will not be
denumerable.   This situation  is similar  to that  pertaining to  the
concept truth, of requiring a metalanguage in order to define truth in
the object language \cite{T44}.

The proof given  above for the uncomputability of $x_0$  is similar to
the  that of  the uncomputability  of the  halting problem  for Turing
machines  \cite{md1958}.  The  concept of  a meta-computer  alluded to
above  indicates that  the proof  of uncomputability  ``relativizes'',
meaning that one can construct harder problems, by allowing a computer
program to call as subroutine an ``oracle'' that solves the above free
choice  problem in  bounded time.   One can  then construct  a G\"odel
sentence for this oracle-enhanced program, which then yields a problem
with  its hardness  shifted  one  level higher  than  the free  choice
problem above.  Problems  on the same level  of uncomputability, i.e.,
uncomputable  problems  which  are Turing-equivalent,  form  a  Turing
degree.   The  process can  be  repeated  to construct  higher  Turing
degrees, i.e.,  the next  higher levels  of more  uncomputable problem
\cite{Ler83}.  It is known  that there are $2^{\aleph_0}$ (uncountably
many) Turing degrees.

We  suggest  that  the  the causal  hierarchy  corresponds  to  Turing
degrees,  but in  inverse ordering,  whereby the  prime intent,  which
arises  beyond  an  infinite  number   of  causal  stages,  is  itself
Turing-computable. But  the descent  of the  will would  correspond to
transition to higher levels of Turing uncomputability, making the free
choice of agents of sufficiently  low FW highly uncomputable.  The act
of FW in general must be a monstrous hypertask, since the lower causal
stages correspond to ever higher orders of uncomputability.

Why should the causal and computational hierarchy correspond with each
other? Here  we will appeal  to a  teleological argument: that  if the
consequences of the causal hierarchy were computable, then there would
have been no need for the  ``brute force'' computation provided by the
physical manifestation of the universe and human agents!
  
Considerable  research has  been devoted  in computability  theory and
mathematical logic to the study of the complicated structure of Turing
degrees.  Perhaps all  of that  may have  a bearing  on the  cognitive
structure of free-willed agents.

\section{Neurological basis for FW \label{sec:noorba}}

The  presence of  the FW  oracle in  the human  brain marks  the basic
difference  between  a  human  agent and  a  robotic  simulation.  The
question of  how the  FW oracle  is embedded in  the brain  and called
forth, is briefly considered here.

A $\mathfrak{F}^\sharp_\infty$  agent is  an infinite  entity, whereas
the  physical  brain of  a  human  being is  finite  in  terms of  its
information storage  and computation  capacity.  Therefore,  if humans
are  $\mathfrak{F}_\infty^\sharp$  agents,  sufficiently  high  levels
$L_j$ cannot have a physical representation, i.e., a neural correlate.

We propose  the following physical  realization of the model.   At the
instance  a human  becomes aware  of  a choice,  she is  instinctively
driven to  enact her ``nature'',  encoded by $P_0$.  At  the selection
window, her ``reason'', which  carries a representation of $P_\infty$,
advices  her to  deviate  $X_0$ towards  $X_\infty$.   This creates  a
potential   tension,  which   may   result  in   a  fleeting   quantum
superposition,  and   may  correlate   with  the   agent  subjectively
experiencing an internal conflict.

\begin{center}
\begin{figure}
\includegraphics[width=9.0cm]{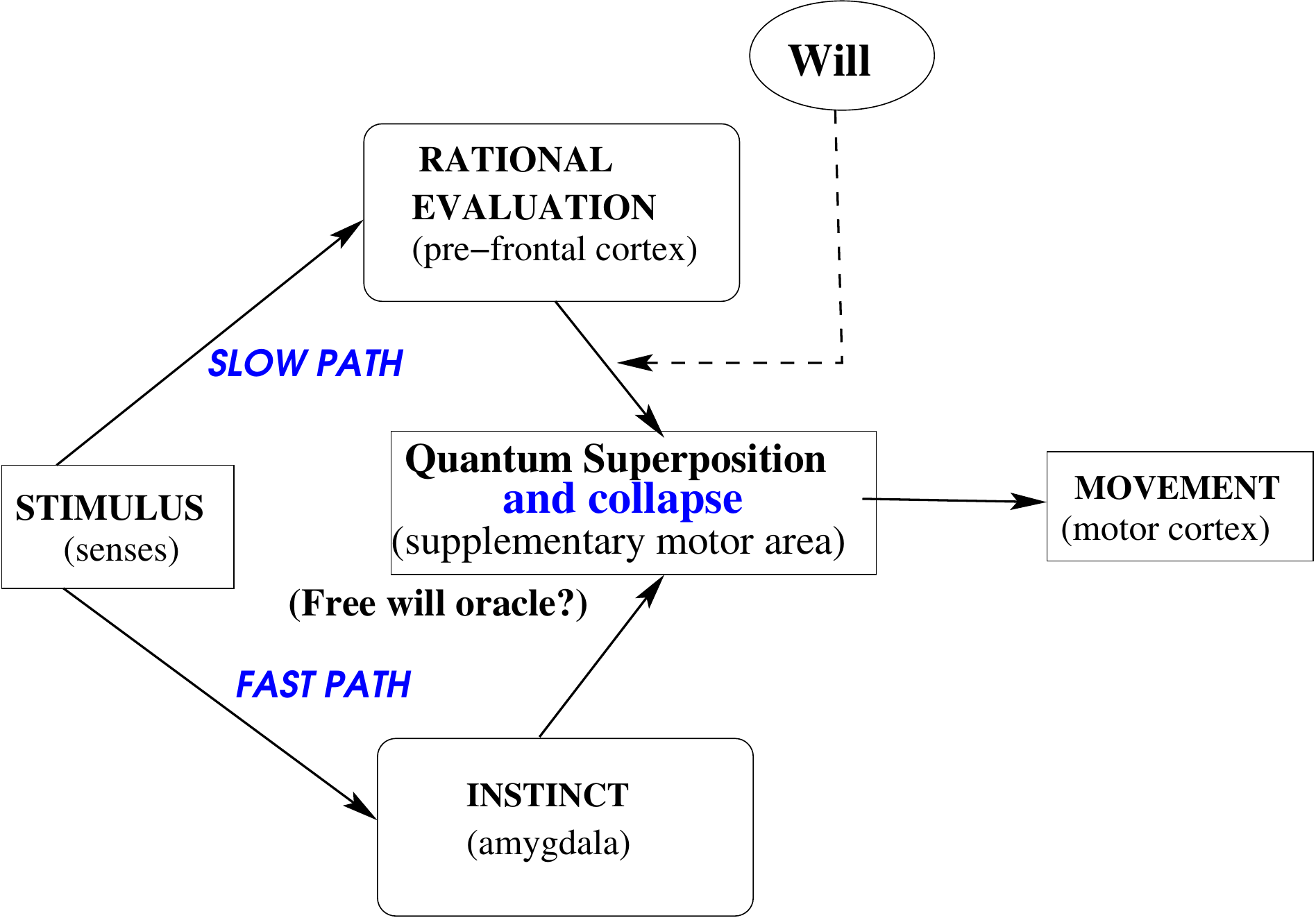}
\caption{Neurobiological flowchart  for free choice by  an agent.  The
  conflict  situation triggers  neural signals  along two  paths-- the
  fast  ``nature''  path  and  the  slow  ``reason''  path--  carrying
  possibly opposing  recommendations for action. If  the neural signal
  in the slow path is weak,  the agent executes the instinctive action
  induced by  the fast path.   But if the signal  in the slow  path is
  sufficiently  strong, and  there  is a  conflict  between $P_0$  and
  $P_\infty$, then a quantum superposition is set up at the FW oracle,
  as the intermediate step for deviating $X_0$ towards $X_\infty$.}
\label{fig:hana}
\end{figure}
\end{center}

Since $P_0$ is  determined by $W_0$, its neural  correlate is expected
to  be well  defined, and  associated  with the  motor cortex.   Since
$P_\infty$  is  largely  computable,  its  neural  correlate  is  also
expected  to  be  well  defined, and  associated  with  the  reasoning
circuits in  the pre-frontal cortex.  In  Figure \ref{fig:hana}, these
two correlates are  represented as the slow and  fast neural pathways,
at whose  confluence the  FW oracle  lies.  

From physically observable data, one may  be able to predict a pattern
of behavior. However, since the  higher levels in the causal hierarchy
are not observable, therefore in any given instance of an agent's free
choice, even the  most detailed neural imaging (say via  fMRI) will be
unable in principle to predict with full certainty what the agent will
select.

The FW mechanism in the  brain of a $\mathfrak{F}_\infty^\sharp$ agent
can never be  modelled as a finite input-output  device.  An important
physical consequence  is that,  the mental  causation that  produces a
deviation from $W_0$  may correspond to deviations  from physical laws
associated with $W_0$,  like energy conservation or the  Second Law of
thermodynamics.  For example,  the  initial voltage  fluctuation in  a
motor  neuron that  initiates the  spontaneous movement  of a  mouse's
whisker, may  be energy-wise  unaccounted physically, even  though the
subsequent nonlinear  amplification of that fluctuation  to a physical
action will certainly be governed by (classical) physics.

%-- supervenience --

%\begin{widetext}
%\end{widetext}

A possible  experimental test  of the model  could aim  to distinguish
willful from  casual choice.  In  the latter case, the  probability of
choice will  reflect $P_0$, which  can be estimated from  the relative
strengths  of signals,  as  picked  up by  fMRI  scans. Under  willful
choice, there  will be a deviation  in the probability of  choice away
from $P_0$ towards $P_\infty$.

\section{Discussions and Conclusions \label{sec:conclu}}

We showed that, although  metaphysical libertarianism is provisionally
protected against the  dilemma of determinism by  the two-stage model,
it  is vulnerable  to  the recursive  version of  the  dilemma.  As  a
defense for libertarian freedom,  we proposed the infinitely recursive
two-stage   scheme.     A   free   agent   is    described   here   by
$\mathfrak{F}_\infty$,       and      free-willed       agents      by
$\mathfrak{F}_\infty^\sharp$.   The  concept  of  FW here  is  of  the
Rationalist-compatibilist kind.

Some   other   issues,  with   ramifications   for  quantum   physics,
neuroscience, mathematics, philosophy, computation theory, are briefly
mentioned below.

There appears to be a parallelism between the will in $\mathfrak{F}_1$
and  hidden  variables  in  ontological models  of  quantum  mechanics
\cite{NS14}.  But  there are two basic  differences: these ontological
models attempt \textit{explain} probabilistic physical laws $W_0$ (and
thus  correspond   to  $\mathfrak{F}_\half$),  whereas  the   will  in
$\mathfrak{F}_1$ may produce  \textit{deviations} from $W_0$.  Second,
the  will  in  $\mathfrak{F}_1$  is  the first  rung  in  an  infinite
hierarchy of higher-order willings,  whereas hidden variable models of
quantum mechanics stop at unit depth.

Neurobiology is  the area  most affected  by our  model, and  also its
possible clearest testing ground.  Experimentally locating the seat of
the  FW oracle  in the  brain, and  working out  how mental  causation
initiates  free-willed   action  in  motor  neurons   will  be  vital.
Experiments that  distinguish willful and casual  choice offer another
window of study the neurobiological circuitry for FW.

This understanding can be  medically useful. By potentially clarifying
the roles neurotransmitters  or receptors play in the  process of free
choice,  it  may  be  able  to suggest  medical  solutions  that  help
encourage  self-controlled  behavior,   by  enhancing  the  ``reason''
pathway, rather  than momentarily  suppressing the  ``nature'' pathway
(Figure \ref{fig:hana}).   Such treatments may be  useful for patients
suffering  from neuropathological  ailments like  obsessive-compulsive
disorder (OCD).

In mathematics  and computation  theory, the relationship  between the
causal hierarchy and  Turing degrees would merit  further study.  This
will help  to elucidate the  scope of AI.  The formalization of  FW as
presented here,  along the  lines of formalization  of the  concept of
truth \cite{T44} would be the first step here.

Our   model   implies   that   high  FW   correlates   with   improved
predictability, i.e., reduced entropy  in $X_0$.  This reduction comes
not by a compensatory increase in entropy elsewhere in the universe as
required by  the Second  Law of  thermodynamics, but  by means  of the
deviation from $W_0$ produced by  mental causation. Departure of $P_j$
from $0_j$ is expressed as a  conflict, and thus entropy, on $L_0$.  A
purely physical means to reduce entropy would be subject to the Second
Law, with  no implications for  the agent's cognitive freedom.  But by
freely reducing this  entropy, the agent is aligning  her $P_j$'s with
$0_j$'s,  and  enhancing  her   freedom.   Thus  concepts  like  moral
responsibility  and  justice  are  helpful  as  props  that  encourage
free-willed behavior, and thereby help reduce disorder, though not
necessarily on the physical level.

This in turn  has implications for evolutionary  biology.  It suggests
that the underlying  force driving evolution was  perhaps not Nature's
quest for  propagating the  species most  successful at  survival, but
instead Nature's  quest for greater freedom.   Darwin-like incremental
evolution evolves quantum  matter ($\mathfrak{F}_\half$ agent) through
lower animals  lacking a CNS  ($\mathfrak{F}_J$ with finite  $J$), and
then through higher animals equipped with a CNS but no cerebral cortex
(free  agents,  $\mathfrak{F}_\infty$),  and from  there,  finally  to
free-willed  agents, $\mathfrak{F}_\infty^\sharp$.   Thus \textit{Homo
  sapiens sapiens} perhaps already  represents the limits of Darwinian
biological  evolution.  The  remaining  evolutionary journey,  towards
greater  freedom of  the will,  is  now ``up  to us''.   It should  be
accomplished through self-determination.

\bibliography{free}

\begin{thebibliography}{11}
\expandafter\ifx\csname natexlab\endcsname\relax\def\natexlab#1{#1}\fi
\expandafter\ifx\csname bibnamefont\endcsname\relax
  \def\bibnamefont#1{#1}\fi
\expandafter\ifx\csname bibfnamefont\endcsname\relax
  \def\bibfnamefont#1{#1}\fi
\expandafter\ifx\csname citenamefont\endcsname\relax
  \def\citenamefont#1{#1}\fi
\expandafter\ifx\csname url\endcsname\relax
  \def\url#1{\texttt{#1}}\fi
\expandafter\ifx\csname urlprefix\endcsname\relax\def\urlprefix{URL }\fi
\providecommand{\bibinfo}[2]{#2}
\providecommand{\eprint}[2][]{\url{#2}}

\bibitem[{\citenamefont{Pink}(2004)}]{Pin04}
\bibinfo{author}{\bibfnamefont{T.}~\bibnamefont{Pink}},
  \emph{\bibinfo{title}{Free Will: A Very Short Introduction}}
  (\bibinfo{publisher}{Oxford}, \bibinfo{year}{2004}).

\bibitem[{\citenamefont{Davis}(1982)}]{md1958}
\bibinfo{author}{\bibfnamefont{M.}~\bibnamefont{Davis}},
  \emph{\bibinfo{title}{Computability and Unsolvability}}
  (\bibinfo{publisher}{Dover}, \bibinfo{year}{1982}).

\bibitem[{\citenamefont{Tarski}(1944)}]{T44}
\bibinfo{author}{\bibfnamefont{A.}~\bibnamefont{Tarski}},
  \bibinfo{journal}{Philosophy and Phenomenological Research}
  \textbf{\bibinfo{volume}{4}} (\bibinfo{year}{1944}).

\bibitem[{\citenamefont{Heisenberg}(2009)}]{H09}
\bibinfo{author}{\bibfnamefont{M.}~\bibnamefont{Heisenberg}},
  \bibinfo{journal}{Nature} \textbf{\bibinfo{volume}{459}},
  \bibinfo{pages}{164} (\bibinfo{year}{2009}).

\bibitem[{\citenamefont{James}(1956)}]{J56}
\bibinfo{author}{\bibfnamefont{W.}~\bibnamefont{James}},
  \emph{\bibinfo{title}{The Dilemma of Determinism: The Will to Believe}}
  (\bibinfo{publisher}{New York, Dover}, \bibinfo{year}{1956}).

\bibitem[{\citenamefont{Nayakar and Srikanth}(2014)}]{NS14}
\bibinfo{author}{\bibfnamefont{C.~M.} \bibnamefont{Nayakar}} \bibnamefont{and}
  \bibinfo{author}{\bibfnamefont{R.}~\bibnamefont{Srikanth}}, in
  \emph{\bibinfo{booktitle}{Nature's Longest Threads: New Frontiers in the
  Mathematics and Physics of Information in Biology}}, edited by
  \bibinfo{editor}{\bibfnamefont{J.}~\bibnamefont{Balakrishnan}}
  \bibnamefont{and} \bibinfo{editor}{\bibfnamefont{B.~V.}
  \bibnamefont{Sreekantan}} (\bibinfo{publisher}{World Scientific Publishing
  Co. Pte. Ltd.}, \bibinfo{year}{2014}).

\bibitem[{\citenamefont{Schopenhauer}(1939)}]{S39}
\bibinfo{author}{\bibfnamefont{A.}~\bibnamefont{Schopenhauer}}
  (\bibinfo{year}{1939}), \bibinfo{note}{from Schopenhauer's essay title2d
  \textit{On the freedom of Human Will}. Translation from German to English in
  `The Philosophy of American History: The Historical Field Theory' (1945) by
  Morris Zucker}.

\bibitem[{N39(1839)}]{N39}
\bibinfo{howpublished}{The Royal Norwegian Academy of Sciences (KNVA)}
  (\bibinfo{year}{1839}), \bibinfo{note}{the predecessor of DNVA posed the
  academic question ``Is it possible to demonstrate human free will and
  self-consciousness?''}.

\bibitem[{\citenamefont{Laroudugoitia}(2009)}]{Lar09}
\bibinfo{author}{\bibfnamefont{J.~P.} \bibnamefont{Laroudugoitia}}, in
  \emph{\bibinfo{booktitle}{Stanford Encyclopedia of Philosophy}}, edited by
  \bibinfo{editor}{\bibfnamefont{E.~N.} \bibnamefont{Zalta}}
  (\bibinfo{year}{2009}),
  \urlprefix\url{plato.stanford.edu/entries/spacetime-supertasks/}.

\bibitem[{\citenamefont{Pe\'cnjak}(2009)}]{Pec09}
\bibinfo{author}{\bibfnamefont{D.}~\bibnamefont{Pe\'cnjak}},
  \bibinfo{journal}{Interdisciplinary Description of Complex Systems}
  \textbf{\bibinfo{volume}{7}}, \bibinfo{pages}{14} (\bibinfo{year}{2009}).

\bibitem[{\citenamefont{Shore and Slaman}(1999)}]{tjump}
\bibinfo{author}{\bibfnamefont{R.~A.} \bibnamefont{Shore}} \bibnamefont{and}
  \bibinfo{author}{\bibfnamefont{T.~A.} \bibnamefont{Slaman}},
  \bibinfo{journal}{Math. Res. Lett.} \textbf{\bibinfo{volume}{6}},
  \bibinfo{pages}{711–722} (\bibinfo{year}{1999}).

\end{thebibliography}

\end{document}